\documentclass{article}
\usepackage{authblk}
\title{The Potsdam astroComb (POCO) Part I: Mode crossing effect in feedback resonators}
\author[1,*]{Daniel Bodenm\"uller}
\author[1]{Kalaga Madhav}
\author[1]{Martin Roth}
\affil[1]{Leibniz-Institut f\"ur Astrophysik Potsdam, An der Sternwarte 16, 14482 Potsdam, Deutschland}
\affil[*]{\textit {dbodenmueller@aip.de}}
\date{}                     
\usepackage[english]{babel}

\usepackage[a4paper,top=2cm,bottom=2cm,left=3cm,right=3cm,marginparwidth=1.75cm]{geometry}

\usepackage{amsmath}
\usepackage{graphicx}
\usepackage[colorlinks=true, allcolors=blue]{hyperref}
\usepackage{mathtools} 
\usepackage{mdframed}
\usepackage{subcaption}
\usepackage[noadjust]{cite} 

\begin{document}
\maketitle

\begin{abstract} 

We investigate theoretically and experimentally the mode interaction in an integrated Silicon Nitride (Si3N4) microring resonator with interferometric coupling realized by a feedback loop as an adjustable optical path path length connecting the ring to the bus waveguide at two coupling sections. From the transmission spectra recorded at different optical path lengths, two resonances, 1596.5~nm and 1570.5~nm, were selected for detailed investigation. Both resonances show the possibility of adjusting the resonance width and depth. However, the transmission spectra around the first resonance also show the effect of mode interaction. This is also well captured in the theoretical model, from which we can derive a coupling rate for the mode interaction of $3.4~\textrm{rad}~\textrm{ns}^{-1}$.
\end{abstract}

\section{Introduction}
Resonators are a crucial component of frequency comb generation, with the resonator properties having a direct influence on the frequency comb characteristics. For the generation of combs with repetition rates (i.e. line spacing) in the GHz range, resonators with a corresponding free spectral range (FSR) - i.e. spatial dimensions in the micrometer range - are required.
The microresonators bring additional advantages, such as a small footprint, reduced cost, an increased quality factor (Q-factor) and mode confinement.
The last two properties help to improve the performance in various applications such as the generation of frequency combs for the calibration of astronomical spectrographs \cite{obrzud_microphotonic_2017,suh_searching_2019} or for data transmission in optical telecommunications networks \cite{jorgensen_petabit-per-second_2022,marin-palomo_microresonator-based_2017}.
The wealth of possible applications, especially as a frequency comb source, motivated the development of microresonators on various material platforms with ultra-high Q-factors of several million \cite{pfeiffer_ultra-smooth_2018,kippenberg_kerr-nonlinearity_2004,fujii_all-precision-machining_2020} up to billions \cite{wu_greater_2020}.
Microring resonators on CMOS-compatible platforms (e.g. silicon or silicon nitride) are of particular interest, as they combine the advantages of scalable manufacturability, reliability and stability with the possibility of integration with other optical and electronic components \cite{pasquazi_micro-combs_2018}.
Relevant design parameters, such as ring circumference and the gap between ring and bus waveguide, are determined prior to fabrication and along with the propagation losses, define the FSR, coupling parameters, resonance extinction and Q-factor.

Extending the architecture appropriately achieves a certain degree of variability, which allows to change the FSR and alter the shape of the resonances even after manufacturing. 
Such architectures include additional active elements which allow for changes of the material refractive index and thus the optical path length, i.e. the phase \cite{chen_compact_2007,zhou_electrically_2007,ou_thermo-optically_2022,yin_high-q-factor_2021}. The index variation rely on the free-carrier plasma dispersion effect (via p-i-n diode) \cite{li_silicon_2007} or on the thermo-optical effect via electrical heater pads \cite{li_bandwidth-tunable_2020}.

However, a change in the refractive index also leads to a change in the mode structures: The mode structure of a ring resonator consists of mode families (or resonance families) each with a defined FSR. A frequency of a resonance of one mode family can be arbitrarily close to a frequency of a resonance of a different mode family. In this case, mode coupling can occur due to material or geometric imperfections. This is referred to as a cancellation of mode degeneracy or mode interaction and can be observed either as a function of the mode number (a static case) or as a function of an actively changing refractive index (a dynamic case) which causes a shift in the distance between two particular resonances.  In a local frequency range, this can lead to strong changes of the FSR of both involved mode families\cite{arianfard_optical_2023} and thus, for example, allow the generation of frequency combs in resonators with normal dispersion \cite{liu_investigation_2014,xue_normal-dispersion_2015} or stimulate the generation of solitons \cite{yang_spatial-mode-interaction-induced_2016} and soliton crystals \cite{cole_soliton_2017,boggio_efficient_2022}.

Here we report the analysis of the phenomenon of avoided mode crossing observed for the first time in a microring resonator with interferometric coupling. We also present a new theoretical description that combines both the effect of interferometric coupling and the effect of avoided mode crossing in one model.

\section{Photonic chip and Experimental setup}

\begin{figure}[h]
\begin{subfigure}{0.5\textwidth}
\includegraphics[width=0.9\linewidth]{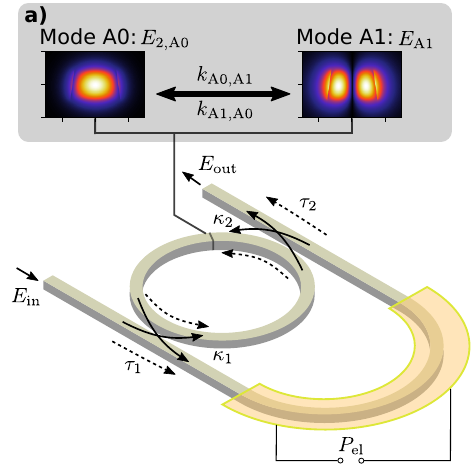}
\caption{}
\label{fig:feedbackResonatorScheme_subimg1}
\end{subfigure}
\begin{subfigure}{0.5\textwidth}
\includegraphics[width=0.9\linewidth]{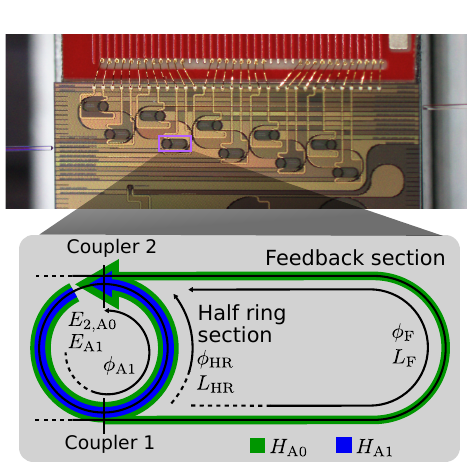}
\caption{}
\label{fig:feedbackResonatorScheme_subimg2}
\end{subfigure}
\caption{ (a) Schematic representation of our microring resonator, which features an interferometric coupling architecture. The two coupling points are denoted by the amplitude coupling coefficients $\kappa_1$,$\kappa_2$ and the amplitude coupler transmissions $\tau_1$,$\tau_2$. Inset a) shows the field distribution over the transverse waveguide section of the first and second TE mode. 
The coupling strength between the associated resonances to these modes is described by the coupling parameters $k_\textrm{A1,A0}$ and $k_\textrm{A0,A1}$. The feedback section is equipped with
an electric heating pad with which the optical path length can be adjusted by applying an electric power $P_\textrm{el}$. (b) A microscope image of the chip with lensed fibers for coupling light in and out. The inset represents the propagation paths of the transfer functions $H_\textrm{A0}$ and $H_\textrm{A1}$ propagating the respective field $E_\textrm{2,A0}$ and $E_\textrm{A1}$ through the resonator.}
\label{fig:feedbackResonatorScheme}
\end{figure}

The design is based on low loss silicon nitride waveguides with a core width of $1.5~\mu m$ in silica fabricated by a PECVD process.
Fig.\ref{fig:feedbackResonatorScheme} shows a schematic view of the resonator together with a photograph of the optical chip.
The waveguide architecture consists of two main elements: the ring in the center with a radius of $132~\mu m$, and the bus waveguide which, after providing a first coupling point, leads back to the ring to provide a second coupler on the opposite side of the ring. The bus waveguide takes on the role of a feedback loop of length $L_\textrm{F}=1.5 \times 132~\mu m$ between two the coupling points. The two coupling points together with the feedback section and the section of the ring form an interferometric coupler.
Electrically driven heating pads on the feedback section allow the phase accumulated along the feedback lengths to be changed by more than $2 \pi$. Thereby it is possible to control the effective power coupled into the ring which also effects the resonance extinction. This also involves small changes in the frequency spacing between resonances directly due to the interferometric effect.\\
The setup for recording the transmission spectra is shown in Figure \ref{fig:aggregated_fig0}a. Light from a tunable infrared laser with defined linear polarization was injected into the chip to excite transverse electric (TE) modes in the bus waveguide while the laser wavelength is swept from 1550 nm to 1630 nm. The outgoing
optical signal was recorded by a photodetector.
For input and output coupling of the chip we used lensed fibers with a focus spot size of $2.5~\mu m$.
In order to obtain an accurate relative frequency axis for the transmission spectra, a fraction of the laser light was simultaneously sent through a fiber-based interferometer.\\

\subsection{Measurement results}
\begin{figure}[!htb]
\centering
\includegraphics[width=1.0\textwidth]{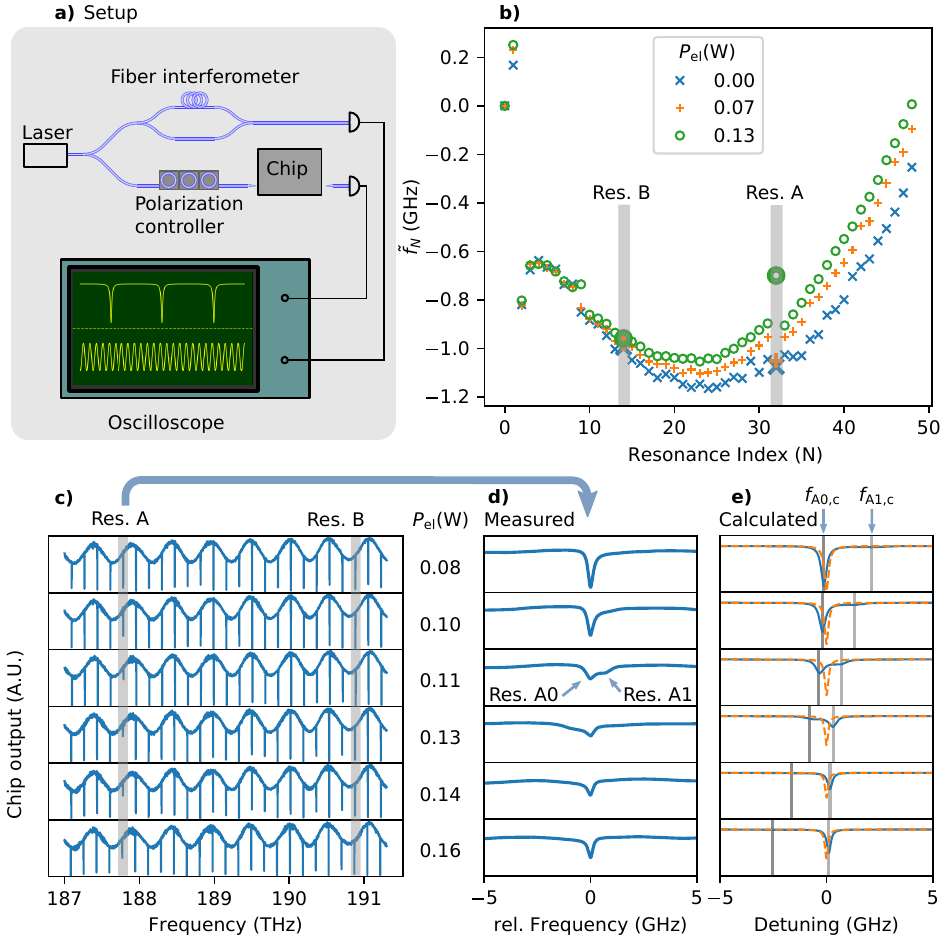}
\caption{(a) Setup for recording the transmission spectra of the resonator with a scanning laser. A fiber interferometer was used to precisely calibrate the frequency axis. (b) shows the deviation of the resonance frequencies from a linear grid given by $\tilde{f_{N}}=f_N-\textrm{FSR}_\textrm{avg} N$ for three electric heating powers where $\textrm{FSR}_\textrm{avg}$ refers to the average free spectral range at $P_\textrm{el}=0$. The graphs show that the FSR increases with the heating power. At $P_\textrm{el}=0.13 \textrm{W}$, the frequency difference of resonance A (i.e., the dominant contribution of A0 or A1) from its neighboring resonances is particularly striking.(c) The transmission spectra for six successive heating powers $P_\textrm{el}$ are shown. While the depth of most resonances does not change significantly, it shows a drastic decrease for resonance A at 0.11 W. A detailed zoom on resonance A in (d) point to the presence of two interacting resonances A0 and A1. The frequency axis is shifted so that 0 refers to the position of the dominant resonance. (e) shows the transmission spectra calculated from the model for comparison. Here the frequency axis is shifted so that 0 indicates the resonance position in the absence of mode interaction. The grey bars indicate the calculated resonance frequencies $f_\textrm{A0,c}$ and $f_\textrm{A1,c}$ as a result of mode coupling. }
\label{fig:aggregated_fig0}
\end{figure}

To highlight the effect of mode coupling, we will examine the transmission spectra of the feedback resonator as a function of the electric heating pad power $P_\textrm{el}$ and select two resonances A and B for a detailed comparison, where resonance B serves as a standard case without mode coupling for the behavior in an interferometrically coupled ring resonator.
Fig.\ref{fig:aggregated_fig0}c shows the transmission spectra for six values of $P_\textrm{el}$ corresponding to different optical path lengths in the feedback section.
The depth of most resonances varies only slightly and decreases with increasing heating power.
One resonance, labeled as resonance A, stands out in particular as its depth changes drastically, indicating a mode interaction. Resonance B in comparison demonstrates the interferometric coupling unaffected by mode interaction.

\begin{figure}[!htb]
\begin{subfigure}{0.5\textwidth}
\includegraphics{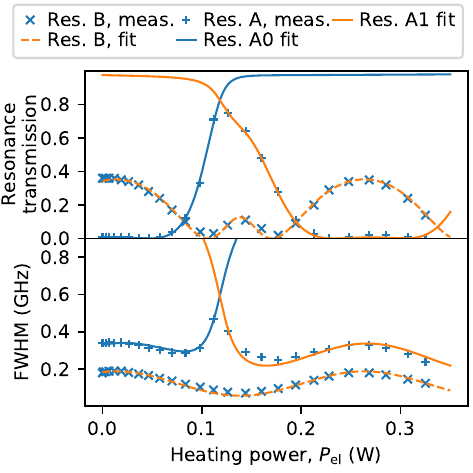}
\caption{}
\label{fig:subim2}
\end{subfigure}
\begin{subfigure}{0.5\textwidth}
\includegraphics{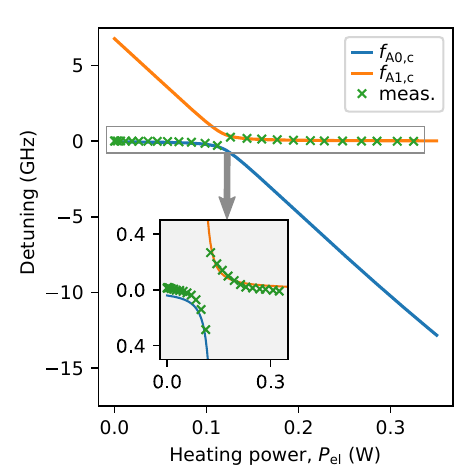}
\caption{}
\label{fig:subim1}
\end{subfigure}
\caption{ (a) The measured resonances transmission and FHWM together with the values calculated from the fitted model. At the crossing point at $P_\textrm{el}=0.11 \textrm{W}$, the dominant contribution switches from resonance A0 to resonance A1. (b) The coupling of two modes (or resonances) A0 and A1 leads to the phenomenon of mode crossing avoidance, which is a shift of the resonance frequencies by $f_\textrm{A0,c}$ and $f_\textrm{A1,c}$ with respect to the resonance frequency in the absence of mode interaction.
The strength of this frequency shift depends on the proximity of the two resonances and therefore on the heating power $P_\textrm{el}$. Values derived from the measurement are shown as crosses and values determined by the model are shown as solid lines. }
\label{fig:image2}
\end{figure}

The measured transmission and full-width-half-maximum (FWHM) for resonance A and B together with the resonance transmissions are shown in Fig.\ref{fig:subim2} over a larger range of $P_\textrm{el}$.
The FWHM of resonance B shows a sinusoidal variation with a maximum of $~190~\textrm{MHz}$ at $P_\textrm{el} = 0$ and a minimum of $~80~\textrm{MHz}$ at $P_\textrm{el} = 0.14~\textrm{W}$. The sinosoidal FWHM variation in interferometrically coupled resonators arises \cite{li_bandwidth-tunable_2020} due to the change in transmission losses experienced by the field within the microring, when passing through the interferometric coupler. This is also accompanied by a change in the resonance transmission, which is observed at the chip output as it changes from 0.38 at $P_\textrm{el}=0$ to nearly zero at $P_\textrm{el}=0.11$ and then to 0.15 at $P_\textrm{el}=0.14$, where the FWHM also reaches its minimum. From the combined observation of the progression of the FWHM and resonance transmission, we can conclude that the coupling condition of resonance B changes from undercoupled to critically coupled to overcoupled at the $P_\textrm{el}$ values mentioned above. Resonance A, on the other hand, shows a different behavior: in Fig.\ref{fig:subim2} the progression of FWHM of resonance A at $P_\textrm{el}$ around 0 and around 0.35~W appears to follow a similar trend to that of resonance B. However, at $P_\textrm{el}$ around 0.11~W, the FHWM increases drastically, indicating a sharp increase in losses. A closer look (see Fig.\ref{fig:aggregated_fig0}d) reveals a peculiar transmission spectrum with a small hump that shifts from the right to the left as the heating power increases. The observed transmission behaviour of resonance A and B will be investigated further and compared to theoretical models we derive in the following sections.

\section{Model for feedback resonator with mode crossing}
In order to understand the observation described above, we derive a model that includes both the inteferometric inteferometric coupling in the feedback resonator and the effect of mode crossing.
The core of our model, similar to equation 1a and 1b in \cite{liu_investigation_2014}, relies on two coupled differential equations describing the time evolution of the intracavity field amplitudes $E_{\textrm{2,A0}}$ and $E_\textrm{A1}$ belonging to the fundamental waveguide mode and to the next higher mode family (primary mode $A0$ and secondary mode $A1$, see inset in Fig.\ref{fig:feedbackResonatorScheme_subimg1}). The amplitudes refer to a specific position in the resonator, namely directly after the second coupler within the microring as shown in Fig.\ref{fig:feedbackResonatorScheme_subimg2}.
To describe how the amplitudes change after each round trip in the resonator, we introduce the transfer functions $H_{\textrm{A0}}$ and $H_\textrm{A1}$. Thus the amplitude $E_{\textrm{A1},N+1}$ after roundtrip $N+1$ results from the amplitude $E_{\textrm{A1},N}$ at round trip $N$  via $E_{\textrm{A1},N+1}=H_\textrm{A1} E_{\textrm{A1},N}$ and correspondingly for $E_{\textrm{2,A0}}$: $E_{\textrm{2,A0},N+1}=H_\textrm{A0} E_{\textrm{2,A0},N}$.
$H_{\textrm{A0}}$ transfers the field amplitude $E_{\textrm{2,A0}}$ from the position directly after the second coupler (see Fig.\ref{fig:feedbackResonatorScheme_subimg2}) through the half-ring section, the first coupler, the feedback section and the second coupler. $H_\textrm{A1}$ transfers the amplitude $E_{\textrm{A1}}$ along the entire circumference of the micro-ring and ignores the feedback section.
With the corresponding time durations $\tilde{t}_{\textrm{r,A0}}$ and $t_\textrm{r,A1}$ for the field transfers $H_{\textrm{A0}}$ and $H_{\textrm{A1}}$ we can formulate a differential equation for $E_{\textrm{2,A0}}$ and $E_{\textrm{A1}}$,  respectively. These are extended by two terms for the mode interaction and a term describing the coupling between the input field $E_\textrm{in}$ and $E_{\textrm{2,A0}}$:

\begin{subequations}
\begin{align}
\tilde{t}_{\textrm{r,A0}} \frac{d E_{\textrm{2,A0}}}{d t} &= \left( H_{\textrm{A0}} (\phi_\textrm{F},\phi_\textrm{HR})-1 \right)  E_{\textrm{2,A0}} + j k_\textrm{A1,A0} E_{A,1} + G (\phi_\textrm{F},\phi_\textrm{HR}) E_\textrm{in} \label{eq:coupledFieldEquation_E12_a}\\
t_\textrm{r,A1} \frac{d E_\textrm{A1}}{d t} &= \left(H_\textrm{A1} (\phi_\textrm{A1})-1\right)  E_\textrm{A1} + j k_\textrm{A0,A1} E_{\textrm{2,A0}} \label{eq:coupledFieldEquation_E12_b}
\end{align}
\end{subequations}

where, $\tilde{t}_{\textrm{r,A0}}$ represents the effective round-trip time (which takes into account the effect of the feedback section) of the field amplitude $E_{\textrm{2,A0}}$, and $t_\textrm{r,A1}$ represents the round-trip time of the field amplitude $E_\textrm{A1}$. $k_\textrm{A1,A0}$ and $k_\textrm{A0,A1}$ refer to the coefficient describing the strength of the coupling between mode $E_{\textrm{2,A0}}$ and $E_\textrm{A1}$. We require for energy conservation, $k=\frac{k_\textrm{A0,A1}}{t_\textrm{r,A1}}=\frac{k_\textrm{A1,A0}}{\tilde{t}_{\textrm{r,A0}}}$. $G (\phi_\textrm{F},\phi_\textrm{HR})$ describes the in coupling via the interferometric section by the input field $E_\textrm{in}$. $\phi_\textrm{F}$, $\phi_\textrm{HR}$ and $\phi_\textrm{A1}$ are the accumulated phases along the feedback section, a half ring section and a full ring revolution, respectively,  for the corresponding mode amplitude and its associated mode family.
$\tilde{t}_{\textrm{r,A0}}$, $t_\textrm{r,A1}$, $H_{\textrm{A0}} (\phi_\textrm{F},\phi_\textrm{HR})$, $H_\textrm{A1} (\phi_\textrm{A1})$ and  $ G (\phi_\textrm{F},\phi_\textrm{HR})$ are expressed as follow:

\begin{equation}
    \label{eq:moreequations}
    \begin{aligned}
    \tilde{t}_{\textrm{r,A0}} &= t_\textrm{r,A0} + (t_\textrm{HR} - t_\textrm{F}) \kappa_1 \kappa_2 a_\textrm{F} a_\textrm{HR} e^{-i(\phi_\textrm{F}+\phi_\textrm{HR})}\\
    H_{\textrm{A0}} (\phi_\textrm{F},\phi_\textrm{HR}) &= \tau_1 \tau_2 a^2_\textrm{HR} e^{-i 2 \phi_\textrm{HR}} - \kappa_1 \kappa_2 a_\textrm{F} a_\textrm{HR} e^{-i(\phi_\textrm{F} + \phi_\textrm{HR})}\\
    H_\textrm{A1} (\phi_\textrm{A1}) &= a_\textrm{A1} e^{-i \phi_\textrm{A1}}\\
    G (\phi_\textrm{F},\phi_\textrm{HR}) &= -i ( \kappa_2 \tau_1 a_\textrm{F} e^{-i \phi_\textrm{F}} + \tau_2 \kappa_1 a_\textrm{HR} e^{-i \phi_\textrm{HR}})    
    \end{aligned}
\end{equation}

where, $t_\textrm{r,A0}$, $t_\textrm{F}$ and $t_\textrm{HR}$ are the group delay (for the primary mode A0) for the lengths of the full ring section $2 L_\textrm{HR}$, the feedback section $L_\textrm{F}$ and the half ring section $L_\textrm{HR}$ which can be expressed with respect to the group index $n_\textrm{g,A0}$ of mode family $A0$ as: $t = \frac{L n_\textrm{g,A0}}{c_0}$ (with $L=2 L_\textrm{HR},L_\textrm{F},L_\textrm{HR}$ for the respective group delay $t=t_\textrm{r,A0},t_\textrm{F},t_\textrm{HR}$). $\tau_1$, $\tau_2$ and $\kappa_1$, $\kappa_2$ refer to the transmission and coupling coefficients of the
first and second coupler, respectively. $a_\textrm{F}$, $a_\textrm{HR}$ and $a_\textrm{A1}$ are the transmission coefficients of the amplitudes for the different waveguide sections given by the propagation losses and can be calculated from the specific waveguide propagation losses via $a=10^{-\alpha L / 20}$ where $\alpha$ represent the respective propagation (power) losses (in dB per unit length) for the various waveguide sections and mode families (with $\alpha=\alpha_\textrm{HR},\alpha_\textrm{F},\alpha_\textrm{A1}$ and $L=2 L_\textrm{HR},L_\textrm{F},L_\textrm{HR}$ for the respective transmission coefficients $a=a_\textrm{F},a_\textrm{HR},a_\textrm{A1}$).\\
While equation \eqref{eq:coupledFieldEquation_E12_a} and \eqref{eq:coupledFieldEquation_E12_b} describe the evolution of the intracavity fields, the field at the output port $E_\textrm{out}$ which is experimentally accessible can be obtained from the $E_{\textrm{2,A0}}$ via:

\begin{equation}
    \label{eq:EoutEquation}
    E_\textrm{out} = (\tau_2 \tau_1 a_\textrm{F} e^{-i \phi_\textrm{F}} - \kappa_1 \kappa_2 a_\textrm{HR} e^{-i \phi_\textrm{HR}}) E_\textrm{in} - i (\tau_2 \kappa_1 a_\textrm{F} e^{-i \phi_\textrm{F}} + \kappa_2 \tau_1 a_\textrm{HR} e^{-i \phi_\textrm{HR}}) a_\textrm{HR} e^{-i \phi_\textrm{HR}} E_{\textrm{2,A0}}
\end{equation}

Note that in our model the amplitude $E_\textrm{A1}$ is coupled neither to the input field $E_\textrm{in}$ nor to the output field $E_\textrm{out}$. This can be justified by the fact that in our measurements we did not directly observe resonances belonging to higher order mode families.
In order to investigate the resonant behaviour of the model we express the phases as a function of the input frequency. Then the phases $\phi_\textrm{F}$, $\phi_\textrm{HR}$ and $\phi_\textrm{A1}$ are not independent anymore and can be expressed as a function of the optical frequency of the input field $\omega$, the optical path lengths of the feedback section $L_\textrm{F,OPL}$ and the optical path lengths of the ring section $2 L_\textrm{HR} n_\textrm{eff}$ :

\begin{subequations}
\begin{align}
    2 \phi_\textrm{HR} &= 2 \pi N + \mu_N + \delta_\textrm{A0}\left(\omega\right) \label{eq:phiHR_a}\\
    \delta_\textrm{A0}\left(\omega\right)&=\frac{\omega-\omega_\textrm{A0}}{\textrm{FSR}_{A0}} \label{eq:phiHR_b}
\end{align}
\end{subequations}

where, $N$ is the mode number, $\delta_\textrm{A0}\left(\omega\right)$ is the detuning from the (uncoupled) resonance frequency $\omega_\textrm{A0}$ and $\textrm{FSR}_{A0}=\frac{c_0}{2 L_\textrm{HR} n_\textrm{g,A0}}$ is the free spectral range for the fundamental mode family in the ring.
The small shift $\mu_N$  is a result of the interferometric effect between the half ring section and feedback section and indicates how far $2 \phi_\textrm{HR}$ has to deviate from $2 \pi N$ to achieve resonant behaviour. It can be approximated as:

\begin{equation}
    \mu_N \approx \frac{2 \kappa_1 \kappa_2 a_\textrm{F} \cos(N \pi )  \sin(\phi_\textrm{F})}{\tau_1 \tau_2 a_\textrm{HR} + \sqrt{(\tau_1 \tau_2 a_\textrm{HR})^2+(\kappa_1 \kappa_2 a_\textrm{F})^2}}
\end{equation}

where $\phi_\textrm{F}$ is expressed as:

\begin{equation}
    \phi_\textrm{F}=\frac{\phi_\textrm{HR}}{L_\textrm{HR}} L_\textrm{F} = \frac{\phi_\textrm{HR} L_\textrm{F,OPL}}{L_\textrm{HR} n_\textrm{eff}}
\end{equation}

For the phase of the entire ring section  $\phi_\textrm{A1}$ of the secondary mode A1 we have:

\begin{equation}
    \phi_\textrm{A1}=2 \pi M + \delta_\textrm{A1}\left(\omega\right)
\end{equation}

where the detuning $\delta_\textrm{A1}\left(\omega\right)$ is related to $\delta_\textrm{A0}$ via:

\begin{equation}
    \label{eq:delta0_relationsship}
    \delta_\textrm{A1} \textrm{FSR}_{A1} = \omega_\textrm{A1} - \omega_\textrm{A0} + \delta_\textrm{A0} \textrm{FSR}_{A0}
\end{equation}

where $\textrm{FSR}_{A1}=\frac{c_0}{2 L_\textrm{HR} n_\textrm{g,A1}}$ is the free spectral range in the ring for mode family A1.

With the equations \eqref{eq:coupledFieldEquation_E12_b} to \eqref{eq:delta0_relationsship}, the model can now be used to determine the transmission $T=|\frac{E_\textrm{out}}{E_\textrm{in}}|^2$ as a function of $L_\textrm{F,OPL}$ - which can be changed via the electrical power applied to the heating pad - and $\delta_\textrm{A0}$ - which is given by the optical input frequency $\omega$. However, it is much more convenient to express characteristic quantities like resonance frequencies and resonances FWHM from our model. Therefore, we consider the homogeneous version of the set of equation \eqref{eq:coupledFieldEquation_E12_a} and \eqref{eq:coupledFieldEquation_E12_b} and obtain the eigenvalues $\lambda_{0,1}$ from which we calculate the coupled mode resonance frequency shifts $\omega_\textrm{A0,c}$ and $\omega_\textrm{A1,c}$ (i.e. the relative frequencies with respect to the uncoupled primary resonance frequency $\omega_\textrm{A0}$) via:

\begin{equation}
    \label{eq:eigenfreqCoupledRes}
    \begin{aligned}
        \omega_\textrm{A0,c} &= 2 \pi f_\textrm{A0,c} = \operatorname{Im} \lambda_0 \\
        \omega_\textrm{A1,c} &= 2 \pi f_\textrm{A1,c} = \operatorname{Im} \lambda_1        
    \end{aligned}
\end{equation}

and the corresponding resonances FWHM via:

\begin{equation}
    \begin{aligned}
        2 \pi \textrm{FWHM}_\textrm{A0,c} &= -2 \operatorname{Re} \lambda_0 \\
        2 \pi \textrm{FWHM}_\textrm{A1,c} &= -2 \operatorname{Re} \lambda_1        
    \end{aligned}
\end{equation}

The stationary case of equation \eqref{eq:coupledFieldEquation_E12_a} and \eqref{eq:coupledFieldEquation_E12_b}, i.e. $\frac{d E_{\textrm{2,A0}}}{d t}=\frac{d E_\textrm{A1}}{d t} =0$, yields for $E_{\textrm{2,A0}}$:

\begin{equation}
    \label{eq:E2Aequation}
    E_{\textrm{2,A0}}(L_\textrm{F,OPL},\delta_\textrm{A0})=-\frac{G (H_\textrm{A1}-1) E_\textrm{in}}{(H_{\textrm{A0}}-1)(H_\textrm{A1}-1) + k_\textrm{A0,A1} k_\textrm{A1,A0}}
\end{equation}

Together with equation \eqref{eq:EoutEquation}, we obtain an expression for the function $E_\textrm{out}(L_\textrm{F,OPL},\delta_\textrm{A0})$, which can be evaluated at $\delta_\textrm{A0}=\frac{\omega_\textrm{A0,c}}{\textrm{FSR}_{A0}}$ and at $\delta_\textrm{A0}=\frac{\omega_{A1,\textrm{c}}}{\textrm{FSR}_{A0}}$ to obtain the resonance extinction, i.e., the depth of the two resonances that occur as a result of the mode crossing effect.
For fitting our experimental data we further extended the model above with a linear model for the optical path length of the feedback section as a function of the applied heating power $P_\textrm{el}$:

\begin{equation}
    \label{eq:linearModelOPL}
        L_\textrm{F,OPL}\left( P_\textrm{el} \right)=m_\textrm{OPL} P_\textrm{el} + n_\textrm{OPL} + L_{F,\textrm{OPL,init}}
\end{equation}

where the inital optical path length is given by $L_{F,\textrm{OPL,init}} = n_\textrm{eff} L_\textrm{F}$, $m_\textrm{OPL}$ is the slope parameter and $n_\textrm{OPL}=L_\textrm{F,OPL}\left( P_\textrm{el}=0\right)$ is an additional offset parameter.
For the frequency separation between the uncoupled primary resonance frequency $\omega_{A,0}$ and the uncoupled secondary resonance frequency $\omega_\textrm{A1}$ we also use a linear model:

\begin{equation}
    \label{eq:linearModelOPL}
        \omega_\textrm{A1} - \omega_\textrm{A0} = m_{MC} (P_\textrm{el} - P_{\textrm{el},0})
\end{equation}

where $P_{\textrm{el},0}$ is the heating power at which the frequency separation is zero.

The complete set of fit parameters consists of
the propagation losses $\alpha_\textrm{F}$ in the feedback section, the propagation losses $\alpha_R$ in the ring section, the power coupling parameters $\theta_{1,2}=\kappa^2_{1,2}=1-\tau^2_{1,2}$ for the two coupling points and the coupling rate $k=\frac{k_\textrm{A0,A1}}{t_\textrm{r,A1}}=\frac{k_\textrm{A1,A0}}{\tilde{t}_{\textrm{r,A0}}}$ in addition to the parameters for the linear functions: $m_\textrm{OPL}$, $n_\textrm{OPL}$, $m_{MC}$, $P_{\textrm{el},0}$.

\section{Fitting procedure and results}

Using the model derived above, we fitted the measured FWHM and transmission of resonance $A$ and resonance $B$ as shown in Fig.\ref{fig:subim2}. In the case of a resonators transmission, the coupling parameter and the losses cannot be unambiguously determined from the resonance transmission or FWHM alone. The model is therefore fitted in such a way that it approximates both the FWHM values and the resonance transmission as well as possible at the same time.
Table \ref{table:1} summarizes the fixed parameters for our model as well as the fit parameters. While the lengths of the various sections were defined in the design, $n_{g,A10}$, $n_{g,A1}$ and $n_\textrm{neff}$ were determined by numerical eigenmode calculations with RSoft\textregistered.

For resonance B the transmission data as well as FWHM data is well reproduced by the model and can be understood by the sinusoidal variation of the power, that is effectively coupled to the ring, and the transmission through the interferometric coupling section. Both are function of optical path length $L_\textrm{F,OPL}$ of the feedback section which in turn depends nearly linear on $P_\textrm{el}$ due to the optothermal effect.
In the case of resonance A, the situation is more complicated as we are in fact observing two resonances (A0 and A1). For reasons of practical feasibility, however, the extracted FWHM and resonance transmission refer to the very resonance that causes the strongest dip, i.e. the dominant resonance, at a given thermal power $P_\textrm{el}$. From this, the model recognizes that for $0 \leq P_\textrm{el} < 0.11$~W the resonance A0 contributes the most to the measured FWHM and transmission, while for $P_\textrm{el} \geq 0.12$~W the resonance A1 dominates.
Around $P_\textrm{el}=0.12$~W where the unperturbed resonance frequency almost coincides, i.e. where $\omega_\textrm{A1} - \omega_\textrm{A0} \approx 0$, the mode interaction is very strong, resulting in higher losses and thus a much wider resonance. Due to much higher losses we also see how the resonance extinction is much weaker (resonance transmission is higher)  since the resonance now is strongly under-coupled.
With the fitted model parameters (table \ref{table:1})  and equation \eqref{eq:eigenfreqCoupledRes} we calculated the coupled mode resonance frequencies shifts $f_\textrm{A0,c}=\omega_\textrm{A0,c}/2\pi$  and $f_\textrm{A1,c}=\omega_\textrm{A1,c}/2\pi$  (or relative eigenfrequencies) as a function of heating power which yields the well known mode crossing avoidance diagram shown in Fig.\ref{fig:subim1}. By subtracting the measured FSR  (averaged over all resonance from the frequency) from the difference between resonance A and its next lower resonance we obtain an estimation for the coupled mode resonance frequencies shifts based on the measurement that can be compared to the frequency shifts from the model in the inset of Fig.\ref{fig:subim1}.

Even though our model was only fitted to the measured resonances transmission and FHWM, the calculated frequency shift matches the measured frequency shifts very well.
Additionally we calculated the transmission spectrum in a small frequency band around the resonance in Fig.\ref{fig:aggregated_fig0}e which demonstrate a excellent agreement with the measured transmission.

\begin{table}[!htb]
\begin{tabular}{ |p{6cm}|p{1.5cm}|p{1.8cm}||p{1.8cm}|p{1.8cm}|  }
 \hline
 \multicolumn{5}{|c|}{Fixed Parameters} \\
 \hline  
 Group index for mode $A0$ & $n_\textrm{g,A0}$ & - & \multicolumn{2}{|c|}{2.104}\\
 Group index for mode $A1$ & $n_{g,A1}$ & - & \multicolumn{2}{|c|}{2.164}\\
 Length half ring section & $L_\textrm{HR}$ & $\mu$m & \multicolumn{2}{|c|}{$0.5 \times 132$}\\
 Initial OPL feedback section & $L_{\textrm{F},\textrm{OPL,init}}$ & $\mu$m  & \multicolumn{2}{|c|}{$1.5 \times 132 \times n_\textrm{eff} $}\\
 Effective refr. index (mode $A0$) & $ n_\textrm{eff}$ &  & \multicolumn{2}{|c|}{1.796}\\ 
 \hline
 \hline
 \multicolumn{5}{|c|}{Fit Parameters} \\
 \hline 
 Name&Symbol&Units&Res. B&Res. $A$\\
 \hline
 Losses in feedback section, mode A0 & $\alpha_\textrm{F}$ & $\textrm{dB}/\textrm{cm}$   & $0.5$ &   $0.7$\\
 Losses in ring section, mode A0 & $\alpha_\textrm{HR}$ & $\textrm{dB}/\textrm{cm}$ & $0.07$ &   $0.27$ \\
 Losses in ring section, mode A1 & $\alpha_\textrm{A1}$ & $\textrm{dB}/\textrm{cm}$ &- &   $2$\\
 Power coupling coefficient Coupler 1 & $\theta_1$ & - & $2.5 \times 10^{-3}$ & $3.2 \times 10^{-3}$ \\
 Power coupling coefficient Coupler 2 & $\theta_2$ & - & $5.8 \times 10^{-4}$ & $6.4 \times 10^{-4}$\\ 
 Mode coupling rate & $k$ & $\textrm{rad}~\textrm{ns}^{-1}$ & - & $3.4$\\
 Slope for $L_\textrm{F,OPL}\left( P_\textrm{el} \right)$ & $m_\textrm{OPL}$ & $\mu\textrm{m}/\textrm{W}$ & 6.3 & 6.3\\
 $L_\textrm{F,OPL}\left( P_\textrm{el}=0 \right)$ & $n_\textrm{OPL}$ &  $\mu\textrm{m}$ & -0.08 & -0.12\\
 Slope for resonance separation & $m_{MC}$ & $\textrm{rad}~\textrm{ns}^{-1}\textrm{W}^{-1}$ & - & $370$\\
 El. power for zero resonance separation & $P_{\textrm{el},0}$ & W & - & $0.118$\\ 
 \hline
\end{tabular}
\caption{Overview of the model parameters. $n_\textrm{g,A0}$, $n_\textrm{g,A1}$ and $n_\textrm{eff}$ were obtained from numerical eigen mode calculations with RSoft\textregistered.}
\label{table:1}
\end{table}

We now discuss two aspects of secondary modes A1 in our architecture: (1) While resonances belonging to the fundamental mode family are easily visible in our transmission spectra, resonances belonging to the higher mode family are generally invisible in our design. There are two reasons for this: (a) the evanescent coupling between the bus waveguide and ring waveguide can only effectively exchange energy between modes with similar propagation constants, i.e. between the same mode families. This has been taken into account in our model where there is no direct coupling term between $E_\textrm{in}$ and $E_\textrm{A1}$. (b) the power transfer between the lensed fiber, which is a single-mode fiber, and the higher waveguide modes in the chip is suppressed because the chip waveguides are equipped with inverse tapers at the input and output sections, which act like filters. Only the mode interaction with the fundamental modes reveals the existence of higher order modes, as our observations show by changes of the resonance transmission and shifts of the resonance frequency. (2) It should be noted that a change in the optical path length of the feedback section can not contribute significantly to a change in the frequency spacing between the primary (A0) and secondary (A1) modes. This can only be achieved by changing the optical path length in the microring. The fact that we nevertheless observe a temperature-dependent mode interaction and thus a change in the resonance distance $\omega_\textrm{A1} - \omega_\textrm{A0}$ can be understood by the fact that the induced temperature change of the feedback path also radiates to the ring path which causes a shift in the resonances of different mode families relative to each other. This can also be seen from the increase in the slope of the resonance frequencies in Fig.\ref{fig:aggregated_fig0}, which points to an increase in the FSR and hence the optical path length in the ring. However, this is not explicitly captured in the model and was instead added manually by the equation \eqref{eq:linearModelOPL}.

\section{Conclusion}

In a ring resonator with interferometric coupling and adjustable path length, we investigated the mode-crossing effect and the interferometric effect by analyzing the transmission spectra around two selected resonances. The interferometric effect shows up in a sinusoidal variation of the FWHM of the resonances in the range from 80~MHz to 190~MHz and a variation of the coupling condition from overcoupled to critically coupled to undercoupled. The mode-crossing effect is revealed by a drastic change of the resonance transmission from 0 to about 0.7, accompanied by a change of the FWHM and a local variation of the resonances FSR of about 350 MHz. 

\section*{Acknowledgements}
This work was supported by Bundesministerium f\"ur Bildung und Forschung under Contract No. 03Z22AN11 Astrophotonics.

\bibliographystyle{unsrt}
\bibliography{citationsFor_feedbackResonatorPaper_20231122}

\end{document}